# Structure and Vertical Modes in Finite 2D Plasma Crystals

K. Qiao and T. W. Hyde[ξ]
*CASPER, One Bear Place 97310,
Baylor University, Waco, TX, USA*

*Abstract*

In this research, formation of finite two-dimensional (2D) plasma crystals was numerically simulated. The structure was investigated for systems with various Debye length and it was found there is a transition from a complete hexagonal structure to a structure with concentric rings on the outer edge and hexagonal lattice in the interior as the Debye length increases. The vertical as well as horizontal oscillation modes thermally excited in the 2D dust coulomb clusters were investigated. The horizontal mode spectra is shown to agree with published results while the vertical mode spectra obtained from numerical simulation and analytical method agree with one another. The frequency of the vertical modes is shown to have a maximum corresponding to the whole system acting as a solid plane. For low frequency modes, the largest amplitude particle motion is concentrated in a few inner rings with the outer rings remaining almost motionless in contrast to the horizontal modes for which the strongest motion of the particles is concentrated in the inner rings at high frequencies.

## I. INTRODUCTION

Complex (dusty) plasma has been a research field of great interest ever since the discovery of plasma crystals in 1994 [1-3], due to its potential application in both astrophysics and semiconductor industry. Most of the plasma crystals investigated in laboratories on earth are two-dimensional (2D) or quasi-2D systems of dust particles levitated near a rf plasma sheath edge above the lower electrode. They are usually levitated by the balance of gravity and electric force in the sheath, and confined on the horizontal direction by either a curved lower electrode or a ring put on the electrode. It has been shown that the potential wells confining the particles on both the vertical [4] and horizontal [5,6] directions are almost parabolic, and widely recognized that the interaction between particles is a Yukawa potential in the form of

$$U(r) = q\exp(-r/\lambda_D)/4\pi\varepsilon_0 r, \quad (1)$$

where $q$ is the dust particle charge, $r$ is the distance between any two particles and $\lambda_D$ is the dust Debye length. Thus a Yukawa system confined by a parabolic potential well on the horizontal direction and a much stronger parabolic potential well on the vertical direction (thus a 2D system will be formed) acts as a good model for a finite 2D plasma crystal.

It has been shown that inside a 3D finite coulomb crystal there is a competition between two forms of ordering, a bcc lattice in the interior, and a concentric shell-structure on the outer edge, by numerical simulation [7]. In this research, the finite 2D Plasma Crystals for varying $\lambda_D$ will be numerically simulated using a Yukawa system of 1000 particles. The dependence of their structures on $\lambda_D$ will be investigated both qualitatively and quantitatively.

Another major topic of research for finite 2D Plasma Crystals (clusters) is the oscillation modes. Horizontal oscillation modes have recently been investigated intensely [5,6]. However, the total mode spectra for 2D dusters should consist of modes not only involving horizontal particle motion but vertical particle motion as well. The dust lattice wave modes created by such vertical particle motion have been investigated for one-dimensional (1D) [8-10] and 2D plasma crystals [11,12] as infinite systems, but the vertical oscillation modes have not yet been examined for finite plasma crystals (clusters). Accordingly, in this research, the vertical as well as horizontal oscillation modes are obtained for particle numbers between $N = 3$ and $N = 150$ employing a box_tree simulation of thermally excited finite 2D plasma crystals. The horizontal mode spectra obtained is then compared with previously published experimental and theoretical results while the vertical mode spectra is analyzed and compared with corresponding analytical results.

## II. NUMERICAL SIMULATION

The formation of finite 2D plasma crystals was simulated using the box_tree code [11-15]. In this research, the interparticle potential is modeled as a Yukawa potential of the form given by Eq. (1). The external confining potential is assumed to be parabolic in 3D,

[ξ] email: Truell_Hyde@Baylor.edu

$$E_{ext}(x,y,z) = \frac{m}{2}\left[\omega^2(x^2+y^2)+\omega_z^2 z^2\right] \qquad (2)$$

where $x, y, z$ are representative particle coordinates, $\omega^2$ is the measure of the strength of the parabolic confinement on the horizontal direction and $\omega_z^2$ on the vertical direction. Despite its simplicity, this potential model captures the basic properties of a multitude of classical systems and serves as an important reference point for more complex confinement situations [16].

Particles are contained in a $10\times10\times10$ cm cubic box, which is large enough that particles, confined by the external potential, always remain within the box for various Debye lengths examined. Particles are assumed to have constant and equal masses of $m_d = 1.74\times10^{-12}$ kg, equal charge of $q = 3.84\times10^{-16}$ C and equal radii of $r_0 = 6.5$ μm. Initially all simulations assume a random distribution of particles subject to the condition that the center of mass of the particle system must be located at the center of the box. Thermal equilibrium for the particle system at a specified temperature is established by allowing collisions of the dust particles with neutral gas particles. Particles are initially given a zero velocity; they obtain higher velocities almost immediately after the start of the simulation due to potential interactions and then cool via collisions with neutrals. Crystal formation occurs less than 5 seconds into the simulation with the particle system continuing to cool slowly thereafter and reaching a thermal equilibrium in approximately 30 seconds. Dimensionless lengths and energies are employed throughout by introducing the units $r_0=(q^2/2\pi\varepsilon m\omega^2)^{1/3}$ and $E_0=(m\omega^2 q^4/32\pi^2\varepsilon^2)^{1/3}$, respectively [16].

## III. CRYSTAL STRUCTURES

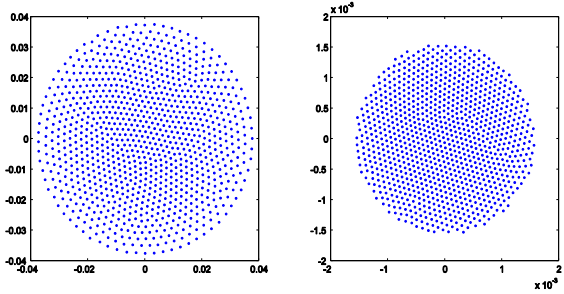

Figure 1. Top view of the finite 2D plasma crystals made up of 1000 particles with (a) $\lambda_D = 7.3123$ and (b) $\lambda_D = .0024$ at T = 10k.

For the first part of this research, systems of particles made up of 1000 particles with various Debye lengths $\lambda_D$ ranging from 10μm to 30mm are simulated at four different temperatures, 1000K, 100K, 10k and 1K. For large $\lambda_D$ (e.g. $\lambda_D = 7.3123$) a 2D lattice with hexagonal lattice in the interior but a few concentric rings on the outer edge is formed (Fig. 1 (a)). As $\lambda_D$ decreases the outer rings disappear and for very small $\lambda_D$ (e.g. $\lambda_D = .0024$) a "pure" hexagonal lattice forms (Fig. 1 (b)). This tendency can be seen more clearly by looking at the radial distribution function of the particles (Fig. 2). Fig. 2 (a) and (b) show the radial density distribution function for $\lambda_D = 7.3123$ and $\lambda_D = .0024$ at T = 10K. There can be seen clearly at least 3 concentric rings on the outer edge of the lattice for $\lambda_D = 7.3123$ but no rings for $\lambda_D = .0024$ at all. On the other hand, Fig 2 (c) and (d) shows the pair correlation functions for these cases and it can be seen that for both large and small Debye length, the lattices are hexagonal in the interior.

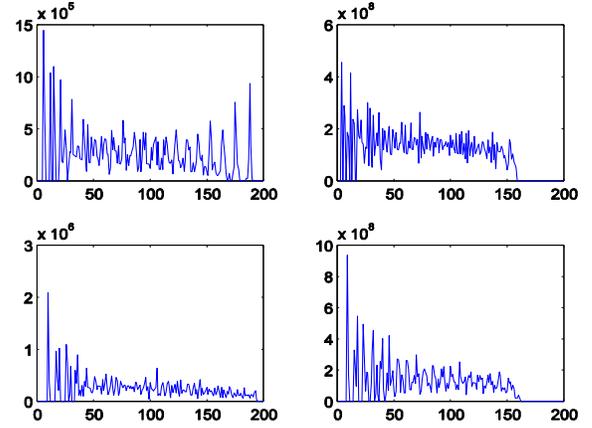

Figure 2. The radial density distribution function of finite 2D plasma crystals for $\lambda_D = 7.3123$ (a) and $\lambda_D = .0024$ (b) and the pair correlation function for $\lambda_D = 7.3123$ (c) and $\lambda_D = .0024$ (d) with N = 1000 at T = 10K.

To quantitatively investigate the effect of $\lambda_D$ on this transition, we have calculated the potential energy of the system versus $\lambda_D$ at four different temperatures. It was found that for the same Debye length, systems with higher temperature have larger potential (Fig.3). This is easy to understand since the higher the temperature is, the larger the average displacements of the particles from their equilibrium positions are. The potential depends on the Debye length much more than it does on temperature. Fig.3 (a) shows the logarithim function of the system potential versus $\lambda_D$. It can be seen that for $\lambda_D < .2437$, the curve is almost linear; for $\lambda_D > .2437$, the curve is linear too but with a smaller slope than for $\lambda_D < .2437$. Thus, it seems like that for the two types of structures, the dependence of potential on $\lambda_D$ is different and there is a transition at around $\lambda_D = .2437$. This tendency can be seen more clearly in Fig.3 (b), which shows the interaction potential rather than the total potential of the system.

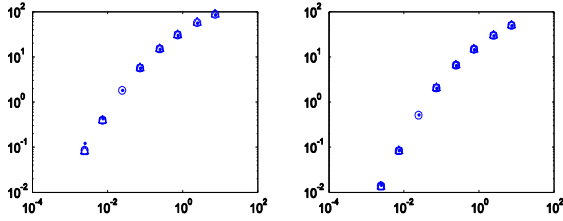

Figure 3. The total (a) and interaction potential (b) of finite 2D plasma crystals made up of 1000 particles as functions of $\lambda_D$ for various temperatures

## IV. VERTICAL OSCILLATION MODES

For the second part of this research, the formation of finite 2D plasma crystals for particle numbers falling between $N = 3$ and $N = 150$ were simulated. The Debye length $\lambda_D = 0.57 mm$. The external potential within the sheath is given by Eq. (2) with $\omega_{xy0}^2 = 2.21$ and $\omega_{z0}^2 = 221.07$. Once the plasma crystal reach the thermal equilibrium, the thermal motion of the dust particles around their equilibrium positions was tracked for 32 seconds with output data files created every 0.04 second yielding a total of 800 data files. These data were used to obtain the normal mode spectra employing the method given by Melzer [6], taking into account not only the horizontal but also the vertical motion of all cluster particles. Additionally, the normal mode spectra were also analytically calculated using the equilibrium positions of the particles. The spectra of the normal modes, including both the horizontal and vertical modes are given in Fig. 4 (d-f). As can be seen, the horizontal mode spectra agrees with published results [5,6] and the simulation results shown in the intensity graphs are in excellent agreement with the theoretical results represented by the solid dots. The spectra for the vertical normal modes are shown in Fig. 4 (g-i). The maximum frequency for a given vertical mode is known as the vertical oscillation frequency. For this case it was found to be $\omega_{z0} = 14.87 s^{-1}$, which corresponds to the oscillation frequency for a vertical oscillation of the entire particle system acting as a solid plane. It is interesting to note that for small numbers of particles, i.e. $N < 47$, the horizontal and vertical spectra appear as two separate branches. As the overall particle number increases ($N \geq 47$), the minimum frequency for the vertical mode decreases until it falls below the maximum frequency for the horizontal modes, and a merging of the two branches occurs.

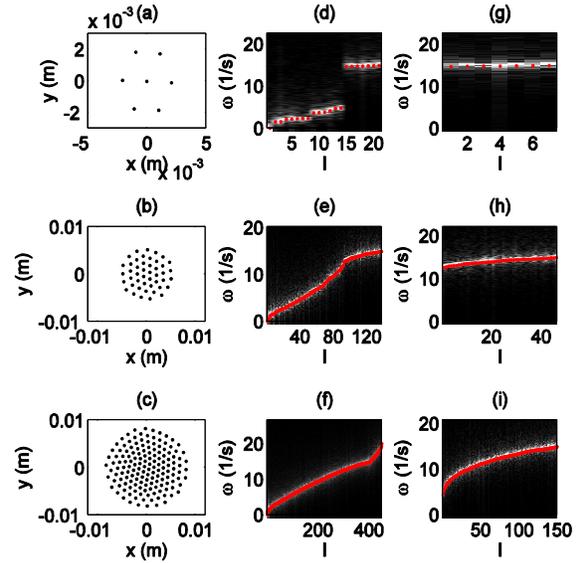

Figure 4. Initial cluster structure for N = 6 (a), 46 (b) and 150 (c). Mode spectra for all normal modes (d-f) and vertical modes only (g-i) for N = 6 (d, g), 46 (e, h) and 150 (f, i).

An detailed investigation of the vertical mode oscillation patterns for the dust clusters mentioned above (particle number $3 < N < 150$) has been conducted by examining the eigenvectors $\vec{e}_{i,l}$. Besides the highest frequency mode mentioned previously, the modes with the second and third highest frequencies correspond to rotational oscillations of the system (again as a solid plane) around two different horizontal axes. For a isotropic system, these two modes are degenerate; in the system modeled above, they have close but slightly different frequencies due to the anisotropic nature of the cluster. For example, when $N = 150$, they have frequencies of $14.7935 s^{-1}$ and $14.7934 s^{-1}$ respectively. These three modes were found to exist for clusters of any particle number.

At the other limit, as shown in Figure 6, for the lowest frequencies (N = 150) peaks and valleys only appear at the center of the cluster while all other areas of the cluster remain flat. Thus, the maximum energy (amplitude) for vertical motion of the particles remains concentrated across a few inner rings with the outer rings remaining almost motionless in the vertical direction. This is contrary to the manner in which horizontal modes act where it is for the highest frequency modes that the particle motion is concentrated within the inner two rings and the outer rings remain more or less motionless.[11] Taken together, for a 2D dust cluster with large particle numbers, horizontal particle motion is distributed throughout the cluster while vertical motion remains concentrated at the cluster's center for low frequencies with exactly the opposite occurring for high frequencies.

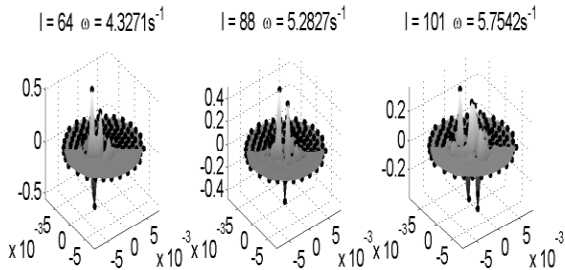

Figure 5. Oscillation patterns for the vertical modes with the lowest frequencies ($N = 150$).

## V. CONCLUSIONS

In this research, formation of finite 2D plasma crystals was numerically simulated as a Yukawa system confined on both the vertical and horizontal direction using box_tree code. It was found that for large $\lambda_D$ a 2D lattice with hexagonal lattice in the interior but a few concentric rings on the outer edge is formed and for very small $\lambda_D$ a "pure" hexagonal lattice forms. This tendency has been seen from the lattice view, the radial density distribution function and the pair correlation function. The transition between the above two type of structures above can be clearly seen from the interaction potential as functions of the Debye length and the critical Debye length was found to be around $\lambda_D = .2437$.

The vertical oscillation modes as well as the horizontal oscillation modes for thermally excited 2D dust coulomb clusters were investigated. The horizontal mode spectra were shown to agree with previously published results while the vertical mode spectra obtained from the box_tree simulation were compared with an analytical method and shown to agree with one another.. For clusters with small particle numbers ($N < 47$), all vertical modes show higher frequencies than corresponding horizontal modes. For larger clusters ($N \geq 47$), the vertical and horizontal modes are co-mingled. It was found that three highest frequency modes exist for all cluster numbers. The highest frequency mode is the mode corresponding to a vertical oscillation of the entire system of particles as a solid plane. This frequency is always equal to $\omega_{z0}$ and is independent of particle number. The modes with the second and third highest frequencies are quasi-degenerate and correspond to rotational oscillations of the system (again as a solid plane) around two different horizontal axes.

At the other limit for the lowest frequency modes, the strongest particle motion within the cluster is concentrated within the first few inner rings with the outermost rings remaining almost motionless. In contrast, the horizontal modes show the strongest particle motion concentrated within the inner rings at their highest frequencies.

## VI. REFERENCES

[1] J. H. Chu and Lin I, "Direct observation of Coulomb crystals and liquids in strongly coupled rf dusty plasmas," Phys. Rev. Lett., vol. 72, pp. 4009, 1994.
[2] Y. Hayashi and K. Tachibana, "Observation of Coulomb-Crystal Formation from Carbon Particles Grown in a Methane Plasma," Jpn. J. Appl. Phys., Vol. 33, pp. L804, 1994.
[3] H. Thomas, G. E. Morfill, V. Demmel, J. Goree, B. Feuerbacher, and D. Möhlmann, "Plasma Crystal: Coulomb Crystallization in a Dusty Plasma," Phys. Rev. Letters, vol. 73, pp. 652, 1994.
[4] E. B. Tomme, D. A. Law, B. M. Annaratone, and J. E. Allen, "Parabolic Plasma Sheath Potentials and their Implications for the Charge on Levitated Dust Particles," Phys. Rev. Lett., vol. 85, pp. 2518, 2000.
[5] V. A. Schweigert and F. M. Peeters, "Spectral properties of classical two-dimensional clusters," Phys. Rev. B, vol. 51, pp. 7700, 1995.
[6] A. Melzer, "Mode spectra of thermally excited two-dimensional dust Coulomb clusters," Phys. Rev. E, vol. 67, pp. 016411, 2003.
[7] Hiroo Totsuji, etc., "Competition between two forms of ordering in finite coulomb clusters," Phys. Rev. Lett. Vol. 88, pp. 125002, 2002.
[8] S. V. Vladimirov, P. V. Shevchenko, and N. F. Cramer, "Vibrational modes in the dust-plasma crystal," Phys. Rev. E, vol. 56, pp. R74, 1997.
[9] T. Misawa, N. Ohno, K. Asano, M. Sawai, S. Takamura, and P. K. Kaw, "Experimental observation of vertically polarized transverse dust-lattice wave propagating in a one-dimensional strongly coupled dust chain," Phys. Rev. Lett., vol. 86, pp. 1219, 2001.
[10] B. Liu, K. Avinash, and J. Goree, "Transverse optical mode in a one-dimensional Yukawa chain," Phys. Rev. Lett. Vol. 91, pp. 255003, 2003.
[11] K. Qiao and T. W. Hyde, "Dispersion properties of the out-of-plane transverse wave in a two-dimensional Coulomb crystal," Physical Review E, vol. 68, pp. 046403, 2003.
[12] K. Qiao and T. W. Hyde, "Structural phase transitions and out-of-plane dust lattice instabilities in vertically confined plasma crystals," Physical Review E, vol. 71, pp. 026406 2005.
[13] D. C. Richardson, "A new tree code method for simulation of planetesimal dynamics," Mon. Not. R. Astron. Soc, vol. 261, pp. 396, 1993.
[14] L. S. Matthews and T. W. Hyde, "Gravitoelectrodynamics in Saturn's F ring: Encounters

with Prometheus and Pandora," J. Phys. A: Math. Gen., vol. 36, pp. 6207, 2003.

[15] J. Vasut and T. Hyde, "Computer simulations of Coulomb crystallization in a dusty plasma," IEEE transactions on plasma science, vol. 29, pp. 231, 2001.

[16] P. Ludwig, S. Kosse, and M. Bonitz, "Structure of spherical three-dimensional Coulomb crystals," Phys. Rev. E, vol. 71, pp. 046403, 2005.